\documentclass{article}
\usepackage{arxiv}
\usepackage[utf8]{inputenc}
\usepackage[T1]{fontenc}
\usepackage{xcolor}
\usepackage[round]{natbib}
\definecolor{blendedblue}{rgb}{0.2, 0.2, 0.6}
\usepackage{graphicx}
\usepackage[bookmarks   = true,
            citecolor   = blendedblue,
            colorlinks  = true,
            linkcolor   = blendedblue,
            urlcolor    = blendedblue,
            citecolor   = blendedblue,
            linktocpage = false,
            hyperindex  = true]{hyperref}
\usepackage{booktabs}
\usepackage{amsfonts}
\usepackage{enumitem}
\usepackage{nicefrac}
\usepackage{amsmath, amssymb, amsthm}
\usepackage{url}
\usepackage{doi}
\usepackage{bm}
\usepackage{fontawesome}
\usepackage{mathtools}
\usepackage{lmodern}
\usepackage{siunitx}
\usepackage{makecell}
\usepackage{booktabs}
\usepackage{etoolbox}
\usepackage{calc}
\usepackage{dsfont}
\usepackage{multirow}
\usepackage{array}
\usepackage{graphicx}
\graphicspath{{}}
\usepackage{todonotes}
\usepackage{changes}

\definecolor{blendedblue}{rgb}{0.2, 0.2, 0.6}
\definecolor{forestgreen(web)}{rgb}{0.13, 0.55, 0.13}
\definecolor{darkorange}{rgb}{1.0, 0.55, 0.0}
\definecolor{emeraldblue}{HTML}{1eb5be}

\newcommand{\norm}[1]{\left\lVert#1\right\rVert}

\usepackage{amsfonts, amsmath, amssymb, enumerate} 

\usepackage[labelsep=period]{caption}

\renewcommand{\headeright}{}

\usepackage{parskip}
\makeatletter 
\renewcommand\@biblabel[1]{#1.} 
\makeatother

\title{
Removal of Multivariate Environmental Influences in Structural Health Monitoring through Conditional Covariances and Supervised Learning} 

\renewcommand{\shorttitle}{Removal of Multivariate Environmental Influences in SHM}

\date{\today 
}

\author{
	\href{https://orcid.org/0000-0003-2256-1127}{\includegraphics[scale=0.06]{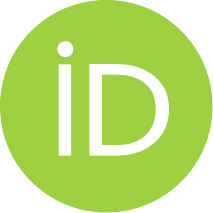}\hspace{1mm}Lizzie Neumann} \\
        Chair of Statistics and Data Science \\
 	Dept.~of Mathematics and Statistics\\
	School of Business, Economics and Social Sciences\\
    Helmut Schmidt University\\
	Hamburg, Germany\\
	\texttt{neumannl@hsu-hh.de} \\
 	\And
	\href{https://orcid.org/0000-0001-7151-8243}{\includegraphics[scale=0.06]{orcid.pdf}\hspace{1mm}Philipp Wittenberg} \\
        Chair of Statistics and Data Science \\
 	Dept.~of Mathematics and Statistics\\
	School of Business, Economics and Social Sciences\\
    Helmut Schmidt University\\
	Hamburg, Germany\\
	\texttt{pwitten@hsu-hh.de} \\
	\And
	\href{https://orcid.org/0000-0001-6777-4746}{\includegraphics[scale=0.06]{orcid.pdf}\hspace{1mm}Jan Gertheiss} \\
        Chair of Statistics and Data Science \\
 	Dept.~of Mathematics and Statistics\\
 	School of Business, Economics and Social Sciences\\
    Helmut Schmidt University\\
	Hamburg, Germany\\
	\texttt{gertheij@hsu-hh.de} \\
}

\begin{document}	
\maketitle

\begin{abstract}
In structural health monitoring (SHM) systems, data is collected from a multitude of sensors measuring, for example, vibration or strain in the structure, along with additional features that capture environmental or operational information. It is well known that changes in the measured sensor outputs do not necessarily originate from structural damage but are often induced by environmental changes. One popular approach to account for these effects is regressing the system outputs on the confounding factors, also known as ``response surface modeling''. Afterward, the predicted values are subtracted from the observed ones to obtain corrected data with the environmental effects (supposedly) removed. However, the evaluation of real-world SHM data shows that environmental conditions may affect not only the expected output values but also higher-order statistical moments, particularly the variances of and the covariances and correlations between the output quantities, such as eigenfrequencies of different modes or strain sensors at different locations. By construction, the (supervised) machine learning techniques commonly used for response surface modeling cannot account for those higher-order effects. 
To address these issues, we present and discuss several approaches for identifying and quantifying multivariate confounding effects on output covariances and correlations: a nonparametric, kernel-based estimator, a random forest, a semiparametric additive model, and a deep learning approach. Furthermore, we show how the resulting conditional covariance matrices can be used in an SHM pipeline. We compare the competing methods on both artificial data and real-world load test data from the Vahrendorfer Stadtweg bridge in Hamburg, Germany, as well as eigenfrequency data from the railway bridge KW51 near Leuven, Belgium.

\end{abstract}

\bigskip

\section{Introduction}
\label{sec_intro}

A primary objective of structural health monitoring (SHM) is to monitor and evaluate engineered systems in civil, mechanical, and aerospace engineering, with the aim of identifying anomalies, such as abrupt structural changes. Typically, sensor measurements are collected, processed, reduced to damage-sensitive features, and further analyzed to distinguish between ``normal'' fluctuations due to environmental and operational variables (EOVs) and abnormal changes due to structural damage. 
The process of removing EOV effects from the data is known as ``data normalization''~\citep{Farrar.Worden_2013}. 
Many suitable methods for data normalization exist~\citep[e.g.,][]{Han.etal_2021,Wang.etal_2022}, but most ignore the environmental effects on the measurement error. Motivated by this, the present paper introduces methodologies to quantify and reduce uncertainty in system outputs, such as sensor measurements or derived damage-sensitive features, in response to varying environmental or operational variables. 

Distinguishing between ``normal'' fluctuations caused by EOVs and abnormal changes resulting from structural damage is a critical yet challenging aspect of moving SHM technology from research to practical field application. The difficulty arises because EOVs can induce changes in measurements of a magnitude similar to that observed with severe structural damage~\citep{Farrar.etal_2000}. This challenge is particularly pronounced in global monitoring approaches, where minor variations in vibration, deflection, or strain are used to detect and localize damage at unknown or hard-to-reach locations. 

Methods for data normalization can be grouped into two categories: ``supervised'' and ``unsupervised''. Supervised methods require the confounding variables to be measured; unsupervised methods do not. 
Supervised methods include, for example, multiple linear regression as used by \cite{Sohn.etal_1999}. \cite{Ni.etal_2005} employ support vector machines, and \cite{Cury.etal_2012} use neural networks. 
Unsupervised methods, on the other hand, include, for example, different forms of principal component analysis (PCA), with, e.g. \cite{Magalhaes.etal_2012} considering the linear version, and \cite{Reynders.etal_2014} using nonlinear kernel PCA. \cite{Worden.etal_2007} used auto-associative neural networks, \cite{Cross.etal_2011} employed cointegration to linearly combine response variables, generating stationary residuals, and \cite{Huang.etal_2023} used kernel canonical correlation analysis to account for nonlinear associations.

In this paper, we focus on settings where EOV measurements are available. All supervised methods mentioned above use forms of mean regression followed by ``residualization'' as it is called in biostatistics~\citep{Zhou.Wright_2015}, which means that (solely) the predicted outputs are used to normalize the data. By doing so, however, covariate effects on the measurement error are ignored. 
It was demonstrated by \cite{Viefhues2020} and \cite{Neumann.etal_2025b}, though, that confounding influences, particularly the temperature, may substantially affect the measurement error, e.g., in terms of the covariances between the measured eigenfrequencies of different modes. To address these issues, \cite{Neumann.etal_2025b} employed a kernel-based estimator of the conditional covariance matrix to develop a conditional version of PCA and a covariate-adjusted version of the Mahalanobis distance, while \cite{Neumann.etal_2026b} reconstructed fully normalized data on the original scale. In both cases, however, only temperature was considered as the confounding variable. Similarly to the conditional PCA from \cite{Neumann.etal_2025b}, \cite{Wei.etal_2024} used ``covariate-regulated'' PCA based on local linear regression to adjust for the effects of the covariates. And \cite{Wittenberg.etal_2026} used covariate-dependent functional principal component analysis (CD-FPCA), which allows the eigenfunctions and eigenvalues used for FPCA to vary with EOVs. As further attempts to go beyond simple mean regression, \cite{Deng.etal_2023} can be mentioned, who considered the relationship between the distributions of different quasi-static responses. 
%

While the literature primarily focuses on temperature as a confounding variable, future long-term monitoring should also evaluate the effects and interactions of multiple confounding EOVs, as discussed in \cite{Rezazadeh.etal_2025} and \cite{Han.etal_2021}. 
In their review, \cite{Keshmiry.etal_2023} identified wind, humidity, boundary conditions, and mass loading as the most frequently examined variables alongside temperature, which was considered in over $50\%$ of the papers examined. 
\cite{Cross.etal_2013} showed that, in addition to temperature and wind, traffic load should not be neglected for lightweight steel structures.  
\cite{Kim.etal_2025} examined a long-span bridge that was primarily affected by wind and traffic. 
\cite{Jesus.etal_2018} examined temperature and traffic load effects on long suspension bridges, and 
\cite{Mu.etal_2020} considered temperature, relative humidity, and traffic volume for their analysis. 
In summary, this means that methods for EOV compensation are needed that can account for (a) multivariate confounding influences, and (b) in a way that goes beyond simple mean regression, particularly considering covariate effects on the output (co-)variances as well. The objective of this paper is to provide such methods.


The remainder of the article is structured as follows. Section~\ref{sec_nonoparm_est_cond_cov} presents and discusses several approaches for obtaining estimates of conditional covariances under the presence of multivariate covariates. 
Section~\ref{sec_SHMmethods} describes the subsequent removal of environmental effects in SHM in the form of a conditional version of the Mahalanobis distance and other approaches based on conditional principal component analysis. 
Section~\ref{sec_validation_of_ana_method} presents a Monte Carlo simulation study to validate and compare the considered approaches for estimating conditional covariances. In Section~\ref{sec_cs}, the proposed methods are applied to two real-world datasets: the load test data of the Vahrendorfer Stadtweg bridge in Hamburg, Germany, and the modal data of the railway bridge KW51 near Leuven, Belgium. Finally, a discussion and concluding remarks are provided in Section~\ref{sec_conclusion}.

\section{Estimation of Conditional Covariances}
\label{sec_nonoparm_est_cond_cov}

In this section, we first describe two existing methods from the literature for estimating conditional covariances: a nonparametric, kernel-based estimate in Subsection~\ref{sec:nwk} and a random forest approach in Subsection~\ref{sec:rf}. Then, we will propose two alternative approaches based on semiparametric additive models and neural networks/deep learning in Subsections~\ref{sec:add} and \ref{sec:nn}, respectively.

\subsection{Revisiting Kernel-based Methods and Random Forest}

The two methods presented first directly target the conditional covariance matrix $\boldsymbol{\Sigma}(\mathbf{z})$ of the system output under consideration, given the covariate vector $\mathbf{z}$, which contains potentially confounding variables; compare \cite{Neumann.etal_2026a}.

\subsubsection{A Nonparametric, Kernel-based Estimate}\label{sec:nwk}

Let $\mathbf{x} = (x_1,\dots, x_p)^\top\in\mathbb{R}^{p}$ be a $p$-dimensional random (output) vector and $\mathbf{z} = (z_1,\dots, z_q)^\top\in\mathbb{R}^{q}$ a $q$-dimensional covariate vector.
Then, the Nadaraya-Watson kernel estimator \citep{Yin.etal_2010,Neumann.etal_2025b} of the conditional covariance is given as
\begin{equation}
    \label{eq_Sdef}
	\hat{\boldsymbol{\Sigma}}(\mathbf{z}; h) = \left\{\sum_{i=1}^n K_h(\norm{\mathbf{z}_i - \mathbf{z}})\left[\mathbf{x}_i - \hat{\mathbf{m}}(\mathbf{z}_i)\right]\left[\mathbf{x}_i - \hat{\mathbf{m}}(\mathbf{z}_i)\right]^\top\right\}\left\{\sum_{i=1}^n K_h(\norm{\mathbf{z}_i - \mathbf{z}})\right\}^{-1},
\end{equation}
where $\mathbf{x}_i = (x_{i1}, \dots, x_{ip})^\top$, $i=1,\ldots,n$, are observations of $\mathbf{x}$, $\mathbf{z}_i= (z_{i1}, \dots, z_{iq})^\top$ is the associated covariate vector, and $\norm{\cdot}$ denotes the a norm, e.g., the euclidean norm; that means, $\norm{\mathbf{z}_i - \mathbf{z}}$ is the (euclidean) distance between $\mathbf{z}_i$ and $\mathbf{z}$.  
$K_h(\cdot)$ is a kernel function with bandwidth $h$, and $\hat{\mathbf{m}}(\mathbf{z}_i)$ is an estimate of the mean of $\mathbf{x}$ at $\mathbf{z}_i$. The conditional mean can also be estimated utilizing a kernel estimator in terms of
\begin{equation}
    \hat{\mathbf{m}}(\mathbf{z}; h) = \left\{\sum_{i=1}^n K_h(\norm{\mathbf{z}_i - \mathbf{z}})\mathbf{x}_i\right\}\left\{\sum_{i=1}^n K_h(\norm{\mathbf{z}_i - \mathbf{z}})\right\}^{-1},
    \label{eq_mdef}
\end{equation}
or any other supervised learning method of choice. For the kernel functions $K_h$, we can, in principle, use any probability density function $K(u)$ that is symmetric around zero and scaled by $K_h(u) = h^{-1}K(u/h)$, where $h > 0$ is the so-called bandwidth. 
The kernel function weights observation $i$ according to the distance between the measured $z_i$ and $z$. Depending on the chosen kernel, the weighting varies slightly. However, because the bandwidth $h$ controls the width of the kernel and thus the smoothness of the estimate, it is more important~\citep{Lopez.Schulz_2016, Gasser.Mueller_1979}. 
In this paper, a so-called Gaussian kernel is used, which equals the normal density with mean zero and the bandwidth playing the role of the standard deviation. 
The bandwidth can be chosen via cross-validation and the following criteria \citep{Petersen.etal_2019}
\begin{align*}
    \hat h_1 &= \underset{h}{\text{argmin}} \sum_{i=1}^{n} \norm{(\textbf{x}_i - \hat{\textbf{m}_i}(\textbf{z}))(\textbf{x}_i - \hat{\textbf{m}_i}(\textbf{z}))^\top - \hat{\boldsymbol{\Sigma}}(\mathbf{z}_i; h)}_\text{F}^2, \\
    \hat h_2 &= \underset{h}{\text{argmin}} \sum_{i=1}^{n} \text{tr}\left(\hat{\boldsymbol{\Sigma}}(\mathbf{z}_i; h)^{\dagger}(\textbf{x}_i - \hat{\textbf{m}_i}(\textbf{z}))(\textbf{x}_i - \hat{\textbf{m}_i}(\textbf{z}))^\top\right),
\end{align*}
where $\norm{\cdot}_\text{F}$ denotes the Frobenius norm, tr$(\cdot)$ the trace, and $^\dagger$ the pseudoinverse. The optimal bandwidth $h$ can then be estimated using the geometric mean $\hat h_\text{opt} = (\hat h_1 \hat h_2)^{1/2}$. 
Either a global bandwidth $h$ can be chosen, or different bandwidths $h_{j,k}$ for each system output pair $k$ and $j$, see \cite{Neumann.etal_2025b} for a more detailed description. 

In principle, this approach is applicable for any number $q$ of covariates, but with increasing $q$ the performance of such nonparametric, kernel-based methods typically deteriorates due to the so-called ``curse of dimensionality''~\citep{Conn.Li_2019}, which means that in the considered $q$-dimensional covariate space even the observation in the training set closest to input $\mathbf{z}$ may be very far away. As a consequence, estimates in many regions of this space will be unreliable.

\subsubsection{Random Forest}\label{sec:rf}

A second approach in the literature for estimating the conditional covariance matrix with multiple covariates is the random forest method proposed by \cite{Alakus.etal_2023}. A random forest is an ensemble of decision trees. 
Decision trees are flowchart-like models used in machine learning to represent possible decisions and their consequences. A tree structure with nodes (root, decision, leaf), from which it derives its name, is used and branches are used to represent the choices, conditions, and outcomes~\citep{deVille_2013}. A random forest combines multiple decision trees to reduce overfitting on the training data. 

The goal of \cite{Alakus.etal_2023} was to train a random forest using the covariates to identify subgroups of observations with distinct covariance matrices, using multiple decision trees created with a specific splitting criterion: 
When training an individual tree to estimate a conditional covariance matrix, the sample covariance matrix of the left node $\mathbf{\Sigma}_\text{L}$ is defined as 
\begin{equation}\label{eq:sampleCov}
    \mathbf{\Sigma}_\text{L} = \frac{1}{n_\text{L} -1} \sum_{i \in \mathcal{I}_\text{L}} (\mathbf{x}_i - \bar{\mathbf{x}}_\text{L})(\mathbf{x}_i - \bar{\mathbf{x}}_\text{L})^\top,
\end{equation}
where $\mathcal{I}_\text{L}$ is the set of the indices of the observations in the left node, $n_\text{L}$ is the left node size, and $\bar{\mathbf{x}}_\text{L} = 1/n_\text{L} \sum_{i \in \mathcal{I}_\text{L}} \mathbf{x}_i$ the (empirical) mean vector of the observations in the left node. The sample covariance matrix of the right node $\mathbf{\Sigma}_\text{R}$ can be estimated equivalently to Eq.~\eqref{eq:sampleCov} with the right node size $n_\text{R}$ and set of indices $\mathcal{I}_\text{R}$. The splitting criterion is given as
\begin{equation}\label{eq:splitcrit}
    \sqrt{n_\text{R} n_\text{R}} \cdot d(\mathbf{\Sigma}_\text{L}, \mathbf{\Sigma}_\text{R}),
\end{equation}
where $d(\mathbf{\Sigma}_\text{L}, \mathbf{\Sigma}_\text{R})$ is the Euclidean distance between the upper triangular part of the matrices. In each iteration of the tree building process, the chosen split variable from the covariates in $\mathbf{z}$ (or, a random subset of variables in $\mathbf{z}$) and the respective cut point maximizes Eq.~(\ref{eq:splitcrit}). This process is repeated until some stopping criterion is met. The random forest then combines multiple trees, each trained on a bootstrap sample of the original data, by taking the average over the resulting outputs. See \cite{Alakus.etal_2023} for further details.

\subsection{Further Approaches using Supervised Learning}\label{sec:rsm}

As an alternative to the estimates presented above, we can make use of the fact that the entry $\sigma_{j,k}(\mathbf{z})$ in row $j$ and column $k$ of $\boldsymbol{\Sigma}(\mathbf{z})$ is the conditional mean of $[x_j - m_j(\mathbf{z})]\cdot[x_k - m_k(\mathbf{z})]$ given $\mathbf{z}$, $j,k=1,\ldots,p$, with $m_j(\mathbf{z})$ being the conditional mean of $x_j$ given $\mathbf{z}$, i.e., the $j$th entry of the conditional mean vector $\mathbf{m}(\mathbf{z})$. All those conditional means, that is, both the entries of $\mathbf{m}(\mathbf{z})$ and $\boldsymbol{\Sigma}(\mathbf{z})$ as functions of $\mathbf{z}$, can be estimated by using a supervised learning method of choice for regression problems. \cite{Wei.etal_2024}, for instance, used a nonparametric kernel-based approach that is closely related to the technique presented in Section~\ref{sec_nonoparm_est_cond_cov}. The main difference is that they used local linear regression instead of a Nadaraya-Watson-type estimator. In this paper, we will generalize this idea by employing methods that are less susceptible to the curse of dimensionality, namely semiparametric, additive models, and deep neural networks.

\subsubsection{Semiparametric, Additive Modeling}\label{sec:add}

The standard approach of response surface modeling uses a regression function $\mathbf{f}_\mathbf{x}(\mathbf{z})$ to account for the association between the system outputs collected in vector $\mathbf{x}$ and the covariate vector $\mathbf{z}$ in terms of
\begin{equation}\label{eq:rsm}
    \mathbf{x} = \mathbf{f}_\mathbf{x}(\mathbf{z}) + \mathbf{\bm{\varepsilon}},
\end{equation}
where $\mathbf{f}_\mathbf{x}(\mathbf{z}) = (f_{1}(\mathbf{z}), \dots, f_{p}(\mathbf{z}))^\top$, and $\mathbf{\bm{\varepsilon}} = (\varepsilon_1, \dots, \varepsilon_p)^\top$ is a zero mean error vector.
The regression function $\mathbf{f}_\mathbf{x}(\mathbf{z})$ corresponds to the conditional mean $\mathbf{m}(\mathbf{z})$. As sketched above, all methods that can be used to estimate an appropriate $\mathbf{f}_\mathbf{x}(\mathbf{z})$ can also be used for estimating conditional covariances. For this purpose, the $p(p+1)/2$-dimensional random vector $\textbf{y}= (y_{1,1},\dots, y_{p,p})^\top\in\mathbb{R}^{p(p+1)/2}$ is created, with entries
\begin{equation}\label{eq_y}
    y_{j,k} = \left[x_{j} - m_{j}(\textbf{z})\right]\cdot\left[x_{k} - m_{k}(\textbf{z})\right], 
\end{equation}
$j \le k$. For the conditional mean of $y_{j,k}$ given $\mathbf{z}$ we have $\text{E}(y_{j,k}|\mathbf{z}) = \sigma_{j,k}(\mathbf{z}) = \sigma_{k,j}(\mathbf{z})$ (the symmetry property of the covariance matrix). Given a training data set $(\mathbf{x}_i, \mathbf{z}_i)$, $i=1,\ldots,n$, and $\mathbf{x}_i = (x_{i1}, \dots, x_{ip})^\top$, $\mathbf{z}_i= (z_{i1}, \dots, z_{iq})^\top$, we can then first estimate the conditional means $\hat{\mathbf{m}}(\mathbf{z}_i) = (\hat{m}_1(\mathbf{z}_i),\ldots,\hat{m}_p(\mathbf{z}_i))^\top$ by using any suitable method for estimating $\mathbf{f}_\mathbf{x}(\mathbf{z})$ from Eq.~\eqref{eq:rsm}. Afterward, those estimated means are plugged in to calculate
\begin{equation}\label{eq_yi}
     y_{i;j,k} = \left[x_{ij} - \hat{m}_{j}(\textbf{z}_i)\right]\cdot\left[x_{ik} - \hat{m}_{k}(\textbf{z}_i)\right] = \left[x_{ij} - \hat{f}_{j}(\textbf{z}_i)\right]\cdot\left[x_{ik} - \hat{f}_{k}(\textbf{z}_i)\right], 
\end{equation}
$i=1,\ldots,n$, $j \le k$. Then, for each combination of $j \le k$, we solve a regression problem with response values $y_{i;j,k}$ and covariate vectors $\mathbf{z}_i$, analogously to finding $\mathbf{f}_\mathbf{x}(\mathbf{z})$ in Eq.~\eqref{eq:rsm}. In this paper, two different methods will be employed: a deep neural network as described in Section~\ref{sec:nn} below, and a semiparametric, additive model.  
In the latter case, the following two models are considered: 

\begin{itemize}
    \item[(i)] \textit{Additive model:} In the purely additive model, the components $f_j(\mathbf{z})$ of the regression function $\mathbf{f}_\mathbf{x}(\mathbf{z})$ in Eq.~\eqref{eq:rsm} have the form $f_j(\mathbf{z}) = \sum_{l=1}^q f_{j,l}(z_l)$. For estimating the potentially nonlinear functions $f_{j,l}$, different methods are available. Here, we follow the popular approach by \cite{Wood_2017}, where each of the component functions is expanded in basis functions in terms of $f_{j,l}(z_l) = \sum_{r=1}^{K_{j,l}} \beta_{j,l,r}b_{j,l,r}(z_l)$ with basis functions $b_{j,l,r}(\cdot)$, basis coefficients $\beta_{j,l,r}$, and $K_{j,l}$ being the respective basis dimension. The models for estimating $\sigma_{j,k}(\mathbf{z})$ through pseudo observations $y_{i;j,k}$ from Eq.~\eqref{eq_yi} are specified analogously. 
    To be sufficiently flexible with respect to the functions that can be fitted, a rich basis is typically used, and a penalty is added to stabilize the resulting estimates and make them interpretable. The most common penalization is in the form of the integrated squared second derivative $\int_\mathcal{D} (f''(x))^2 dx$, where $\mathcal{D}$ is the domain of the function $f$. See \cite{Wood_2017} for further details.\smallskip 

    \item[(ii)] \textit{Additive model with interactions:} The second class of models allows two-way interactions between the covariates, which means that the terms $f_{j,l}(z_l)$ from above are replaced by $f_{j;l,g}(z_l,z_g)$. In a model like this, the influence of one covariate, say $z_l$, is allowed to change with the other(s). The interaction can be modeled using a so-called tensor product basis; see, e.g., \cite{Wood_2017} for further details.  
    In theory, interaction terms beyond two-way interactions can also be considered. However, this can be challenging due to the required computing resources and the substantial amount of data needed to reliably estimate the corresponding functions. Furthermore, the resulting functions are hardly interpretable, while two-way interactions can still be visualized as surfaces in $\mathbb{R}^3$.
\end{itemize}

\subsubsection{Neural Networks}\label{sec:nn}

\begin{figure}[h]
    \centering
    \includegraphics[width = .75\linewidth]{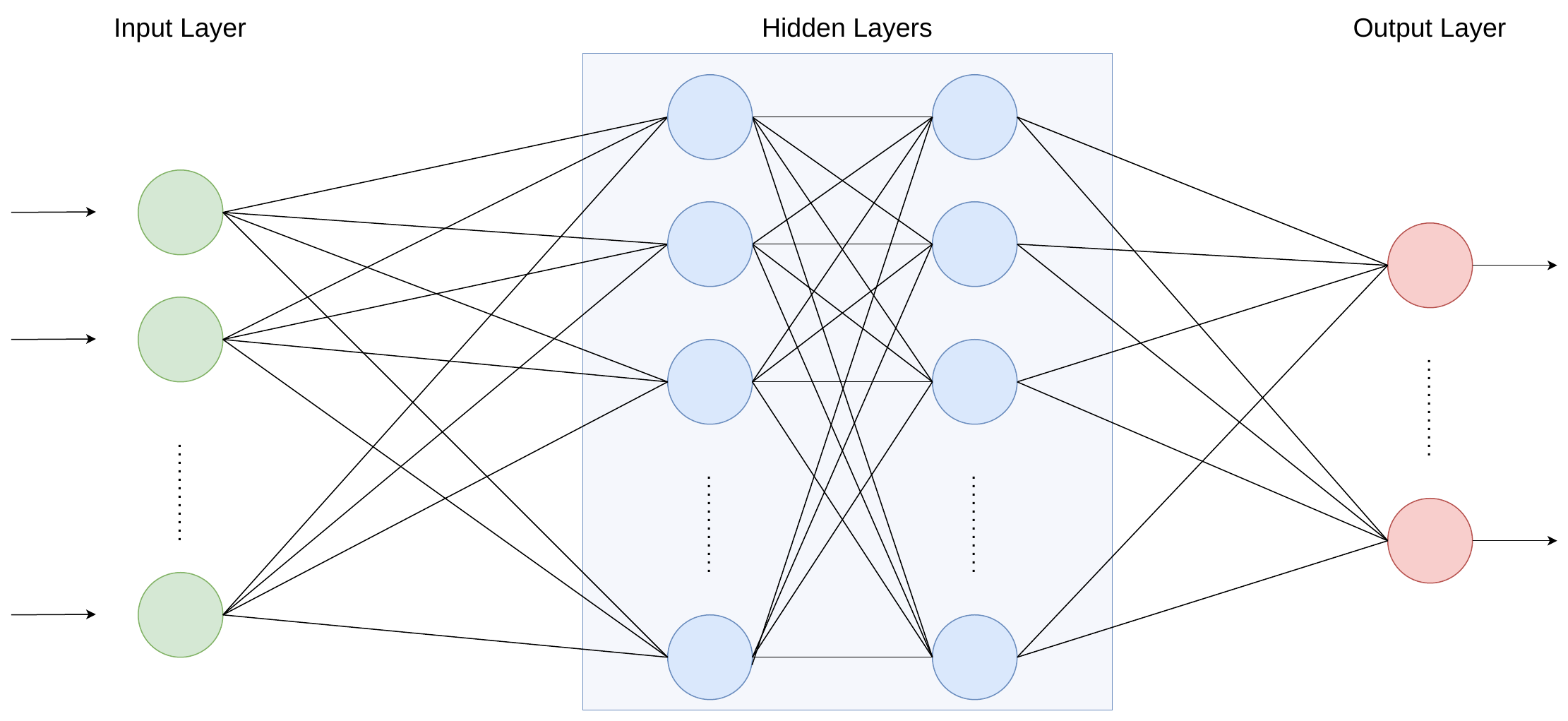}
    \caption{Schematic drawing of a neural network with multiple inputs (green), two hidden layers (blue), and a multi-dimensional output layer (red).}
    \label{fig:nn}
\end{figure}

An artificial neural network is a computational model frequently used in machine learning and artificial intelligence. The structure is inspired by biological neural networks. Different groups of nodes (or artificial neurons) are connected to one another and organized in different layers: the input, hidden, and output layers. A neural network with at least two hidden layers is often called a deep neural network~\citep{Bishop_2006}. The evaluation can be interpreted as forward propagation. 
Backpropagation enables efficient training of multi-layer neural networks by propagating error gradients backward through the layers~\citep{Rumelhart.etal_1986}. 
A schematic drawing of a neural network with two hidden layers is given in Figure~\ref{fig:nn}.

To estimate the conditional covariance of $\textbf{x}$ at $\textbf{z}$ using neural networks, the random vector $\textbf{y}$ from Eq.~\eqref{eq_y} is used, and the conditional mean of $\textbf{y}$ at $\textbf{z}$ is estimated. The neural network is constructed as follows 
\begin{equation}\label{eq_y_nn}
   \hat{\textbf{y}} = f_m(\textbf{W}_m f_{m-1}(\dots f_1(\textbf{W}_1 f_0(\textbf{W}_0\mathbf{z}))),
\end{equation}
with activation functions $f_r(\mathbf{u}) = f(\mathbf{u}) = \max(0,\mathbf{u})$ working component-wise. Matrix $\textbf{W}_0$ contains the weights of the input layer, and $\textbf{W}_r$, $r=1,2,\dots,m$, the weights of the $m$ hidden layers. Backpropagation is performed using the mean square error (MSE) as the loss function and the Adam optimizer. Here, the neural networks are implemented in \texttt{Julia}~\citep{Bezanson.etal_2017} using the \texttt{Flux} library~\citep{Innes_2018,Flux.jl_2018}. Using a neural network as described above, each entry of the conditional covariance matrix can be estimated with basically any number $q$ of covariates that work as the inputs.

\subsubsection{Additional Regularization regarding the Covariance Matrix}\label{sec:reg}

Particularly if using the interpretation of the covariance as a conditional expectation and supervised learning techniques for regression problems, such as the additive model from Section~\ref{sec:add} or the neural network from Section~\ref{sec:nn}, it sometimes happens that the estimated covariance matrix is not positive (semi-)definite for some $\mathbf{z}$. This is problematic, for example, if the inverse $\boldsymbol{\Sigma}(\mathbf{z})^{-1}$ is needed at some point in the subsequent monitoring pipeline (compare, e.g., Section~\ref{sec:CMD} below). An issue that is similar to ill-posed discriminant analysis when a (relatively large) covariance matrix needs to be estimated in a relatively small data group. In the latter case, a popular method to mitigate this problem is regularization as proposed by~\cite{Friedman_1989}. An analogous approach can be chosen here. Specifically, the estimated covariance matrix $\hat{\boldsymbol{\Sigma}}(\mathbf{z})$ from above is replaced by~\citep{Friedman_1989}
\begin{equation}\label{reg_Sig}
    \bar{\boldsymbol{\Sigma}}(\mathbf{z}) = (1-\rho) \hat{\boldsymbol{\Sigma}}(\mathbf{z}) + \frac{\rho}{p} \text{tr}(\hat{\boldsymbol{\Sigma}}(\mathbf{z}))\mathbf{I},
\end{equation}
where $\text{tr}(\cdot)$ denotes the trace of a matrix, i.e., the sum of the diagonal entries, and $\mathbf{I}$ is the identity matrix of suitable dimension; $\rho$ is a tuning parameter that controls shrinkage towards a multiple of the identity matrix. Typically, a small value of $\rho$ is sufficient to stabilize the resulting estimate of $\boldsymbol{\Sigma}(z)$ and produce a valid covariance matrix.

\section{Removal of Environmental Effects in SHM}
\label{sec_SHMmethods}

Once the conditional covariance matrix of the system outputs, given the environmental factors, has been estimated, this matrix, together with the conditional mean, can be used to remove environmental effects from the data. In this section, we will describe several ways and options for doing that. One option is the \textit{conditional Mahalanobis distance}.  

\subsection{Conditional Mahalanobis Distance}\label{sec:CMD}

In statistical process monitoring and SHM, the squared Mahalanobis distance (MD) is a standard tool to quantify the significance of changes in observed outputs. In the univariate case, this is done through the number of standard deviations from a reference value squared, i.e., the MD evaluates the deviation of the measured system output from a reference value, e.g., the covariate-adjusted mean value from training. To obtain a standardized version of these deviations in the case of multivariate outputs, the covariance matrix of the system outputs is required, and usually, a constant covariance matrix is assumed. To fully account for confounding influences, however, the potential effects of EOVs on the covariance should also be considered. For doing that, the conditional covariance can be utilized to turn the standard version into the conditional Mahalanobis distance (CMD); see \cite{Neumann.etal_2025b}, who introduced the CMD for one-dimensional covariates. Extending this approach to multivariate confounders, however, is straightforward in terms of
\begin{equation}
    d_\text{CMD}^2(\textbf{z}) = 
    (\mathbf{x} - \mathbf{m}(\textbf{z}))^\top 
    \boldsymbol{\Sigma}(\textbf{z})^{-1} 
    (\mathbf{x} - \mathbf{m}(\textbf{z})),
    \label{eq_mahalanobis}
\end{equation}
where $\mathbf{x} \in \mathbb{R}^{p}$ is a $p$-dimensional output vector and
$\mathbf{m}(\mathbf{z})\in \mathbb{R}^{p}$ is the conditional mean vector given the multivariate $\mathbf{z}$, which can be estimated through Eq.~\eqref{eq_mdef}. The conditional covariance matrix
$\boldsymbol{\Sigma}(\mathbf{z})\in \mathbb{R}^{p\times p}$ can be estimated using the kernel-based Nadaraya-Watson estimator in Eq.~\eqref{eq_Sdef}, the random forest approach, Eq.~\eqref{eq:sampleCov}, or a supervised learning approach as described in Section~\ref{sec:rsm}, for example, the additive model with or without interactions, Eq.~\eqref{eq_yi}, or the neural network, Eq.~\eqref{eq_y_nn}. Both the mean and the covariance are evaluated for the specific state of the confounder vector $\mathbf{z}$. Thus, the conditional Mahalanobis distance accounts for the confounder $\mathbf{z}$ by incorporating it into both the mean and covariance functions. By using the CMD, the diagnostic takes into account, for instance, that system outputs can exhibit greater variation at certain temperature values than at others.

For online monitoring based on \eqref{eq_mahalanobis}, we follow the standard terminology of statistical process monitoring and define Phase~I as the ``in-control'' reference period used for model fitting (i.e., estimation of the conditional mean function $\mathbf{m}(\textbf{z})$ and the conditional covariance matrix $\boldsymbol{\Sigma}(\textbf{z})$) and control chart calibration. It is assumed to contain only in-control observations, meaning the structural system is stable, and the observed variation can be explained by common-cause variation. Phase~II denotes the subsequent monitoring period. Parts of Phase~II may still be in-control, whereas out-of-control observations correspond to structural changes, damage, or other special causes that are not part of the reference condition. A multivariate Shewhart-type control chart is obtained by plotting $d_\text{CMD}^2(\mathbf z_i)$ sequentially over time and comparing it with an upper control limit $h$. If $d_\text{CMD}^2(\mathbf z_i) > h,$ the observation is flagged as an alarm. Instead of relying on a parametric reference distribution for the monitoring statistic, the control limit can be estimated from the empirical reference distribution of the Phase~I values. Specifically, for a nominal exceedance probability $\alpha$, we set
\begin{equation*}
    h = \widehat Q_{1-\alpha}
    \left(
    d_\text{CMD}^2(\mathbf z_i),\ i \in \text{Phase I}
    \right),
\end{equation*}    
where $\widehat Q_{1-\alpha}$ denotes the empirical $(1-\alpha)$-quantile. This empirical-reference construction avoids imposing a normal or chi-square reference distribution on the plotted statistic \citep{Willemain.Runger_1996}. The resulting chart is deliberately straightforward, as it requires only a reference model for the conditional mean and covariance and an empirical threshold for the resulting scalar monitoring statistic. Its application will be demonstrated later in Section \ref{sec_cs}. If the Phase~II state is known, threshold crossings in in-control parts of Phase~II are interpreted as false-positive alarm time points, whereas threshold crossings in labeled out-of-control segments are interpreted as true-positive alarm time points. Because SHM monitoring statistics are often temporally dependent, these counts are interpreted as alarm time-point counts rather than as counts of independent alarm events.

\subsection{Conditional Principal Component Analysis and Feature Reconstruction}

A second, often used method for removing environmental influences in SHM is principal component analysis (PCA); see, e.g., \cite{Flexa.etal_2019}, \cite{Reynders.etal_2014}. 
Furthermore, it can be used for dimension reduction, anomaly detection, and feature extraction \citep{Tibaduiza.etal_2016, Zhu.etal_2019}. Some PCA-based features are robust to environmental changes \citep{Kumar.etal_2020}, whereas the first principal component(s) mainly account(s) for EOVs \citep{Cross.etal_2012}. This also makes PCA an attractive approach for unsupervised data normalization.
PCA works by rotating the coordinate system to align with the primary directions of variation in the data. This allows users to select specific variations for further analysis, and/or reconstruct a refined signal by back-transforming the chosen components into the physical domain. 
However, a key challenge when using PCA for unsupervised data normalization is that the number of components to omit is unknown a priori. If too few are omitted, confounding effects will remain. On the other hand, if too many components are omitted, damage-relevant information may be removed as well. Furthermore, it is questionable whether this implicit approach can completely eliminate the influence of EOVs \citep{Neumann.etal_2025b, Neumann.etal_2026b}. Therefore, accounting for confounding effects (e.g., due to changing temperatures) more explicitly may be preferable. The conditional PCA \citep{Neumann.etal_2025b}, also called covariate-regulated PCA \citep{Wei.etal_2024}, is an approach to do so. It uses the eigendecomposition of the conditional covariance matrix
\begin{equation}\label{eq:cPCA}
    \mathbf{\Sigma}(\textbf{z}) = \textbf{A}(\textbf{z})\mathbf{\Lambda}(\textbf{z})\textbf{A}(\textbf{z})^\top,
\end{equation}
with the conditional eigenvalues $\lambda_1(\textbf{z}), \lambda_2(\textbf{z}),\dots,\lambda_p(\textbf{z})$, and the conditional principal components $\textbf{a}_1(\textbf{z}), \textbf{a}_2(\textbf{z}), \dots, \textbf{a}_p(\textbf{z})$. These can then be used for feature extraction \citep{Neumann.etal_2025b}
\begin{equation}\label{eq:scores}
    \textbf{s}_i = (\textbf{x}_i - \hat{\textbf{m}}(\textbf{z}_i))^\top \textbf{A}(\textbf{z}_i)(\mathbf{\Lambda}(\textbf{z}_i))^{-1/2},
\end{equation}
with $(\mathbf{\Lambda}(\textbf{z}_i))^{-1/2} = \text{diag}(\lambda^{-1/2}_1(\textbf{z}_i), \lambda^{-1/2}_2(\textbf{z}_i),\dots,\lambda^{-1/2}_p(\textbf{z}_i))$, and $\hat{\textbf{m}}(\textbf{z}_i)$ being an estimate of the conditional mean of $\textbf{x}$ at $\textbf{z}_i$. If the conditional mean and covariances are estimated on ``in-control'' data, the scores~\eqref{eq:scores} obtained on ``in-control'' data are uncorrelated quantities with mean zero for any given $\textbf{z}$-value. If all conditional scores are used, monitoring the squared norm $\sum_{k=1}^{p}s_{ik}^2$ is equivalent to monitoring the squared CMD in Eq.~\eqref{eq_mahalanobis}; hence, the CMD chart can be interpreted as a Hotelling-type chart \citep{Hotelling_1947} in the conditional score space. Alternatively, selected scores can be monitored individually or jointly. Individual score components may be monitored using standard univariate control charts, such as Shewhart and EWMA charts \citep{Lucas.Saccucci_1990}, or CUSUM charts \citep{Page_1954}. This allows distinguishing between a global deviation from the reference condition and changes that occur primarily in specific conditional score directions. Joint monitoring of selected or all conditional scores is also possible, for example, using a multivariate Hotelling-type chart. If the objective is to detect persistent small shifts rather than isolated large deviations, the score vector can instead be monitored over time using a MEWMA chart \citep{Lowry.etal_1992}.


As an alternative to using the scores~\eqref{eq:scores} directly for monitoring, they can also be used to reconstruct the output data on the original scale, with the confounding effects removed. Specifically, we have \citep{Neumann.etal_2026b}
\begin{equation}\label{eq_reconFeat}
    \tilde{\textbf{x}}_i= \bar{\mathbf{x}} + \textbf{A} (\bm{\Lambda})^{1/2} \textbf{s}_i^\top,
\end{equation}
with $\bar{\mathbf{x}}$ the marginal mean of $\mathbf{x}$, $(\bm{\Lambda})^{1/2} = \text{diag}(\lambda_1^{1/2},\ldots,\lambda_p^{1/2})$ the eigenvalues, and $\textbf{A} = [\mathbf{a}_1 \ldots \mathbf{a}_p]$ the principal components from partial PCA. The latter refers to applying standard PCA to the residual data, which means subtracting the (estimated) conditional mean at the confounder values from the observed measurements. The advantage of output reconstruction~\eqref{eq_reconFeat} is that this step can also be interpreted as a preprocessing, which does not induce a specific method for damage or anomaly detection. Consequently, the reconstructed data with confounding influences removed can be used as input to any subsequent SHM pipeline of choice.

\section{Monte Carlo Simulation Study}
\label{sec_validation_of_ana_method}

A Monte Carlo simulation study was conducted to validate and compare the methods described in Section~\ref{sec_nonoparm_est_cond_cov}.
Five different methods were considered: the Nadaraya-Watson kernel estimator from Section~\ref{sec:nwk}, the random forest from Section~\ref{sec:rf}, and three further supervised learning approaches from Section~\ref{sec:rsm}: the additive model without (i) and with interactions (ii) from Section~\ref{sec:add}, and the neural network from Section~\ref{sec:nn}.

\subsection{Experimental Setup}

\begin{figure}[h]
    \centering
    \includegraphics[width=0.32\linewidth]{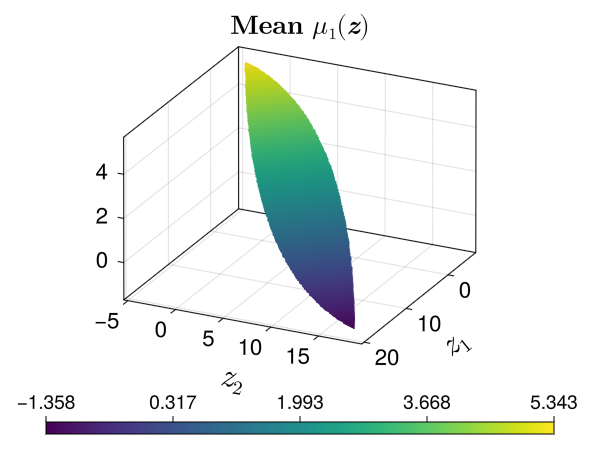}
    \includegraphics[width=0.32\linewidth]{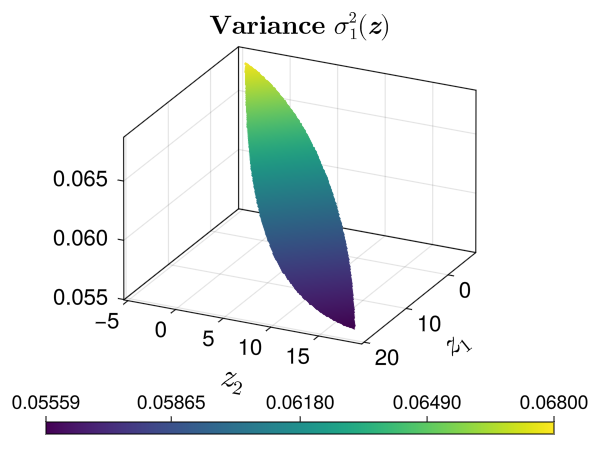}
    \includegraphics[width=0.32\linewidth]{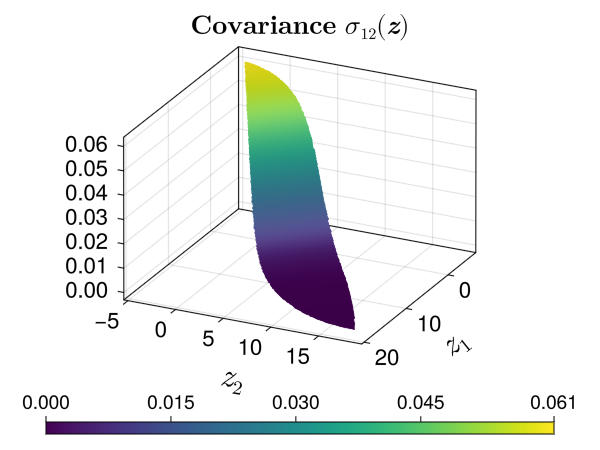}\\
    \includegraphics[width=0.32\linewidth]{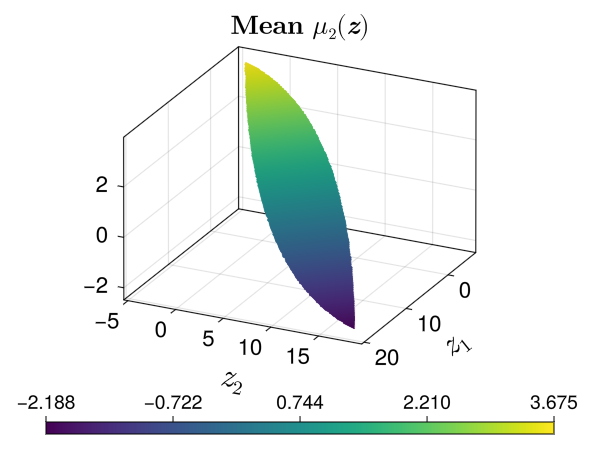} 
    \includegraphics[width=0.32\linewidth]{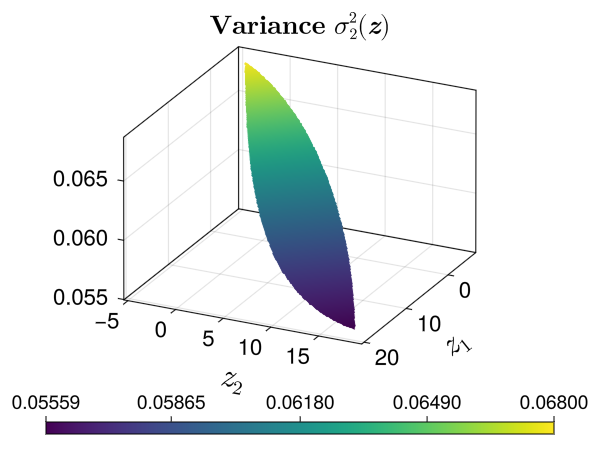}
    \includegraphics[width=0.32\linewidth]{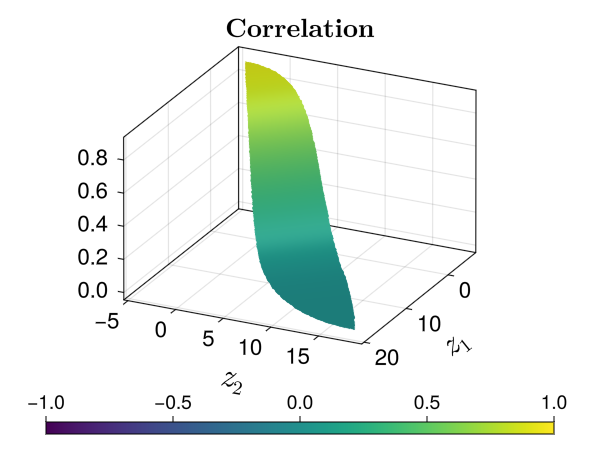}
    \caption{Simulation study setup. Right: conditional mean $\mu_1(\mathbf{z})$ (top) and $\mu_2(\mathbf{z})$ (bottom). Middle: conditional variance $\sigma^2_1(\mathbf{z})$ (top) and $\sigma^2_2(\mathbf{z})$ (bottom). Right: conditional covariance $\sigma_{1,2}(\mathbf{z})$ (top) and resulting correlation (bottom).}
    \label{fig:setup}
\end{figure}

The setup of the Monte Carlo simulation study is similar to that of \cite{Neumann.etal_2026a}. 
We consider a latent two-dimensional normal $\mathbf{x}_t = (x_{1t}, x_{2t})^\top$ with a conditional mean $\mathbf{m}(\mathbf{z}) = (\mu_1(\mathbf{z}),$ $\mu_2(\mathbf{z}))^\top$ and a conditional covariance matrix 
$$\Sigma(\mathbf{z}) = 
\begin{bmatrix}
    \sigma_1^2(\mathbf{z}) & \sigma_{12}(\mathbf{z})\\
    \sigma_{12}(\mathbf{z}) & \sigma_2^2(\mathbf{z})
\end{bmatrix},$$ 
uncorrelated across $t \in \mathbb{Z}$. The observable system outputs are $y_{jt} = x_{jt} + \delta_{jt}$ with zero-mean, homoscedastic AR(1) error process $\delta_{jt}$, and independent across $j$. This AR(1) error process induces temporal correlation in the $y_{jt}$. The resulting covariance $\text{Cov}(y_{1t},y_{2t})$ then is $\sigma_{12}(\mathbf{z}_t)$ from above, and $\text{Var}(y_{jt}) = \sigma_j^2(\mathbf{z}_t) + \nu_j^2$, with $\nu_j^2 = \text{Var}(\delta_{jt})$, where $\nu_1^2 = 0.02$ and $\nu_2^2 = 0.017$ is chosen here.
The mean, (co)variance, and correlation functions are shown in Figure~\ref{fig:setup}, where only the areas with covariate data are depicted.
The latter are inspired by typical temperature profiles and are modeled similarly to those in \cite{Neumann.etal_2026a}.
Specifically, to simulate typical daily and annual patterns for two ``temperature sensors'' the following setup is used:
%
\begin{equation*}
    \mathbf{z}_d(\eta) = \left(
        \begin{array}{ll}
        z_{1,d}(\eta) \\
        z_{2,d}(\eta)
        \end{array}\right)
        = \left(
        \begin{array}{ll}
            8\text{sin}((d-141)2\pi/365) -\zeta_d \text{sin}(\pi\eta/12+0.3) + 5.5\\
            7.5\text{sin}((d-150)2\pi/365) -\zeta_d \text{sin}(\pi\eta/12+0.3) + 5.5
        \end{array}\right),
\end{equation*}
with day $d \in \{1,2,\ldots,365\}$, hour $\eta \in (0,24)$, and $\zeta_d \sim U(a,b)$, where $U(a,b)$ denotes the uniform distribution over the interval $(a,b)$. Different intervals $(a,b)$ were used to ensure more variation on warmer days and less on colder days, as well as different variations for the two different temperatures. 
%

\subsection{Results}

The conditional mean and covariance are estimated using the five approaches from above in the following way. The Nadaraya-Watson kernel estimators in Eqs.~\eqref{eq_Sdef} and ~\eqref{eq_mdef} are used with the Euclidean norm and a global bandwidth of $1.5$, respectively. 
For the random forest approach, the \texttt{R} package \texttt{randomForest}~\citep{Liaw.Wiener_2002} is used for the conditional mean, and the \texttt{CovRegRF}~\citep{Alakus.etal_2023} is used with $100$ trees to estimate the conditional covariance.
For the additive models, the \texttt{R} package \texttt{mgcv} \citep{Wood_2011,Wood_2017} is used, with penalized (cubic) regression splines as regression functions, and, for the additive model with interactions, a joint, bivariate regression function in the form of a tensor product is additionally included. For model fitting, a quasi-likelihood approach is used with an identity link and constant variance.
The neural networks are implemented using \texttt{Flux} \texttt{Julia} \citep{Innes_2018,Flux.jl_2018} as described in Section~\ref{sec:nn}. Two hidden layers are used: the first with size $5q= 10$ and the second with $2q= 4$. For each of the two conditional means, the variance, and the covariance, a separate neural network is trained.
Before calculating $\textbf{y}$ from Eq.~\eqref{eq_y}, the data was standardized with the marginal standard deviation because of the different value ranges of the measurement data.
 
\begin{figure}[h]
    \centering
    \includegraphics[width=0.32\linewidth]{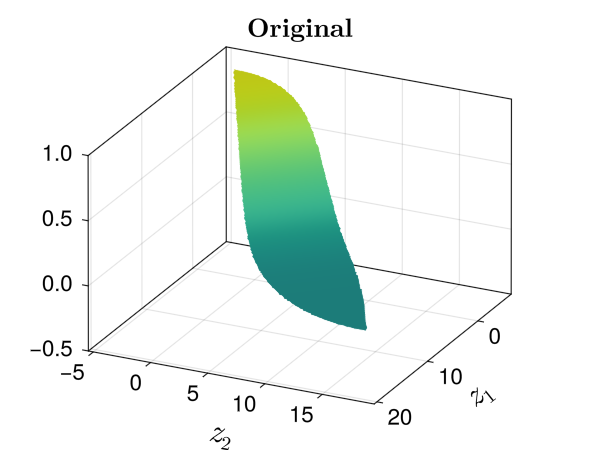}
    \includegraphics[width=0.32\linewidth]{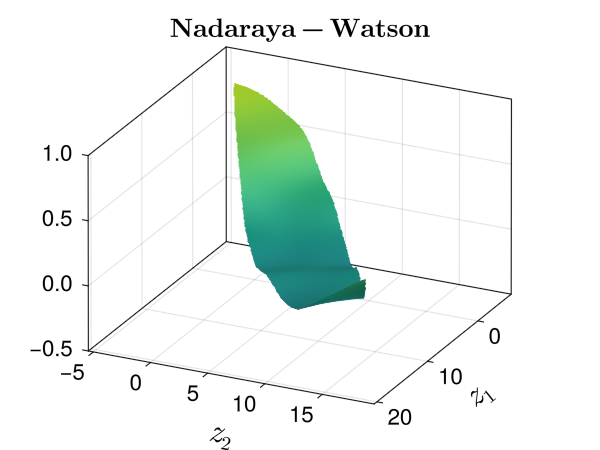}
    \includegraphics[width=0.32\linewidth]{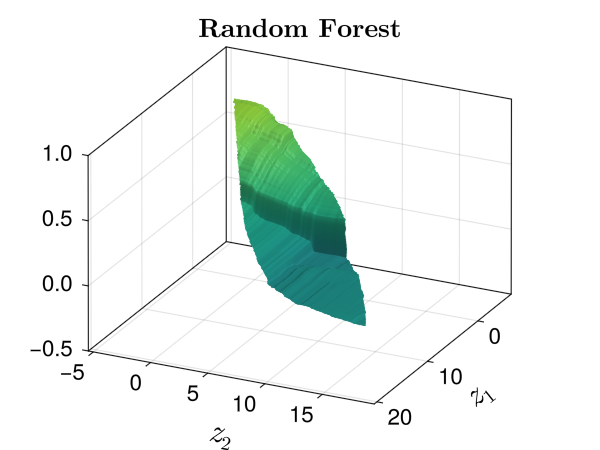}\\
    \includegraphics[width=0.32\linewidth]{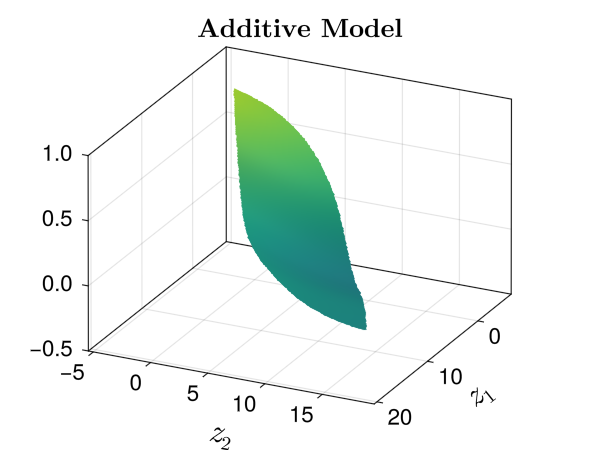}
    \includegraphics[width=0.32\linewidth]{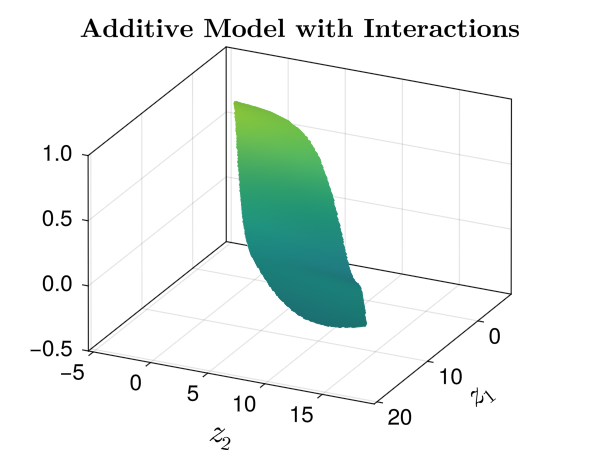}
    \includegraphics[width=0.32\linewidth]{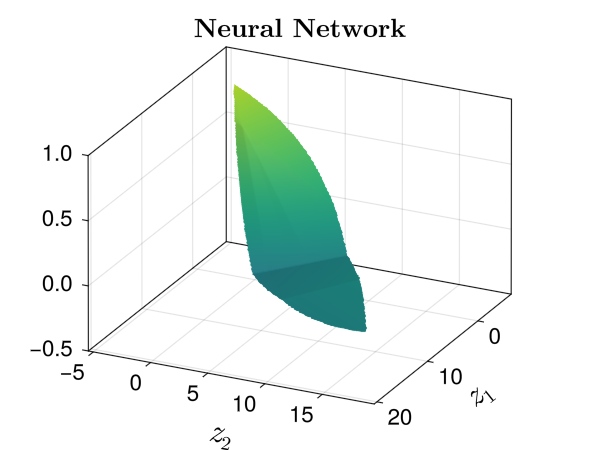} \\
    \includegraphics[width=.75\linewidth]{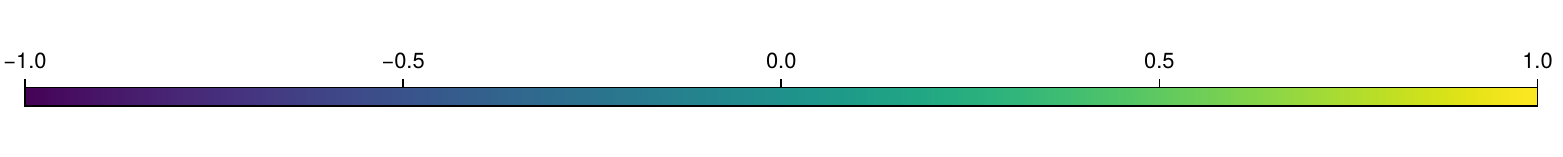}
    \caption{Conditional correlation as a function of the two covariates $z_1$ and $z_2$, and the corresponding estimates on a single dataset. Top: original/true correlation function (left), Nadaraya-Watson (middle), random forest (right). Bottom: additive model (left), additive model with interactions (middle), neural network (right).}
    \label{fig:results}
\end{figure}

For illustration, the estimated conditional correlations from the five approaches are shown for one simulated dataset in Figure~\ref{fig:results}. The true correlation function (top-left) is also shown (again) for comparison. The conditional correlations estimated through the Nadaraya-Watson-type estimator (middle) and the random forest (right) are shown in the top row. The results for the other models are shown in the bottom row, with the additive model on the left and the interaction model in the middle. The neural network estimate is found on the right. Each method is evaluated on a grid of covariate values, but, as in Figure~\ref{fig:setup}, only the areas with covariate data are shown here. 

Each estimated conditional correlation resembles the original correlation in shape. With the exception of the random forest, where the piecewise-constant trees are still visible, all estimates present smooth surfaces. The main difference is observed in the warm-temperature region, where fewer data points are available for training, which leads to greater uncertainty and less stable estimates; this is particularly evident for the Nadaraya-Watson kernel estimator.

\begin{figure}[h]
    \centering
    \includegraphics[width = \linewidth]{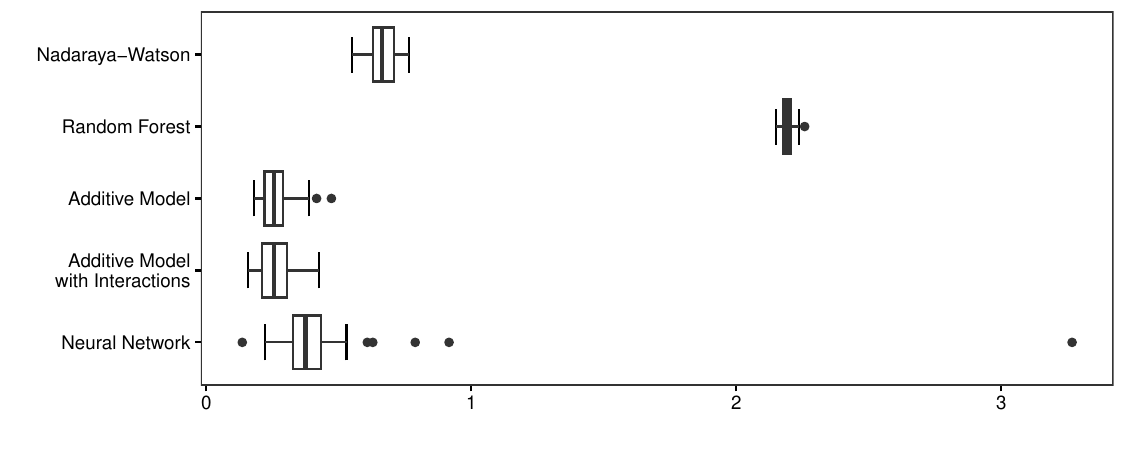}
    \caption{Root mean square error (RMSE) of the conditional covariance estimates for the five considered methods over 50 simulation runs.}
    \label{fig:ss_mse}
\end{figure}

In Figure~\ref{fig:ss_mse}, the root mean square error (RMSE) of the conditional covariances $\hat\sigma_{1,2}(\mathbf{z})$ is shown. Specifically, the conditional covariance was estimated 50 times (i.e., on 50 simulated datasets) by each method, and the MSE was then calculated per method and dataset as the average squared error over the ``observed'' covariate data. The results over the 50 simulation runs are summarized by the boxplots in Figure~\ref{fig:ss_mse}.
The additive models performed best, followed by the neural network and the Nadaraya-Watson kernel estimator. The random forest performed distinctly worse than all the other methods. The random forest's bad performance, however, could be due to the fact that the true covariance function is very smooth (see Figure~\ref{fig:setup}, top-right), which is hard to fit by combining piecewise constant trees (compare Figure~\ref{fig:results}, top-right).

\section{Real-World Case Studies}\label{sec_cs}

In this section, the proposed methods are applied to two real-world data sets. Measurements of strain sensors and eigenfrequencies are analyzed using different covariates (material, steel surface, and asphalt temperature and relative humidity) to show the flexibility of the methods.  

\subsection{Vahrendorfer Stadtweg Bridge}\label{sec_vs}

\begin{figure}[h]
    \centering
    \includegraphics[height = 4.2cm]{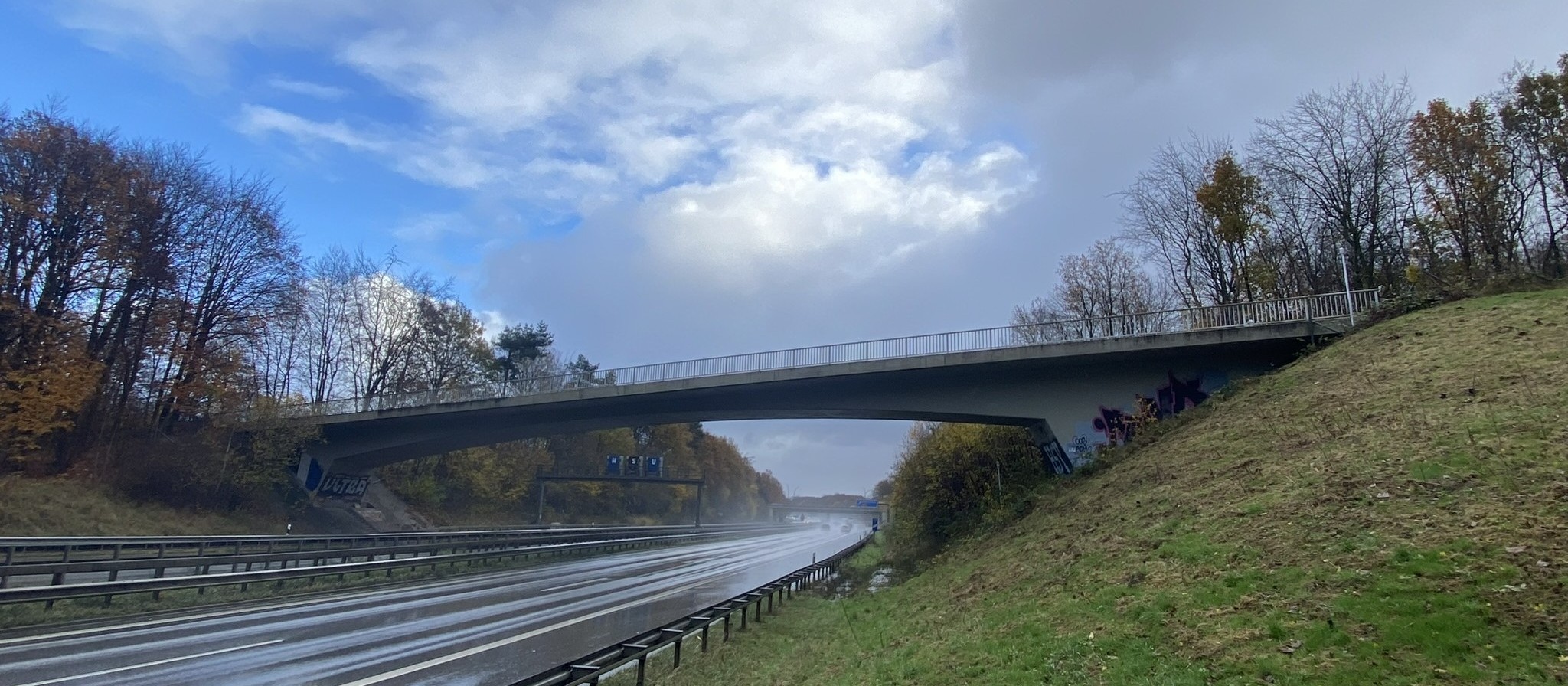}
    \includegraphics[height = 4.2cm]{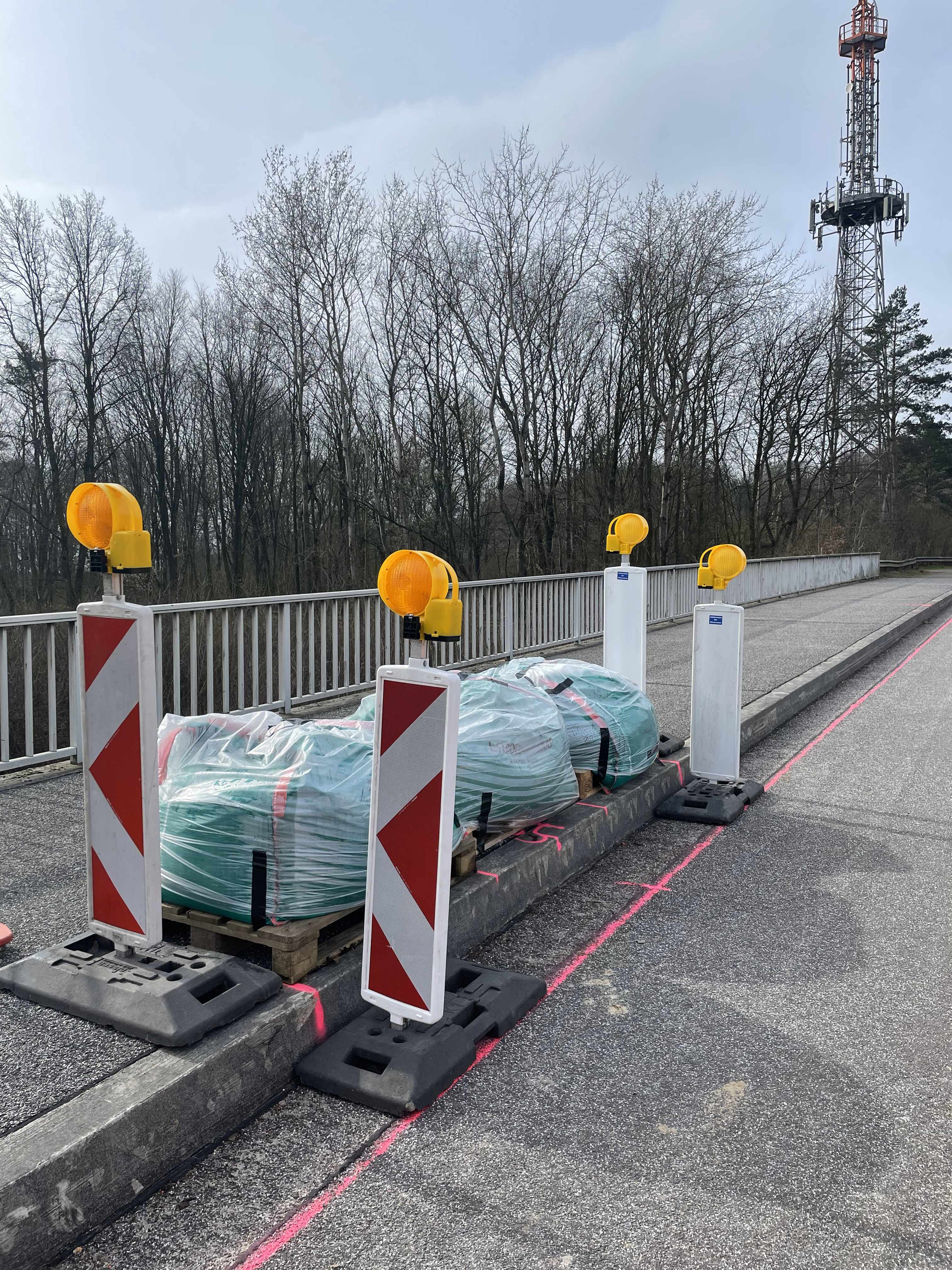}\\
    \includegraphics[width = \textwidth]{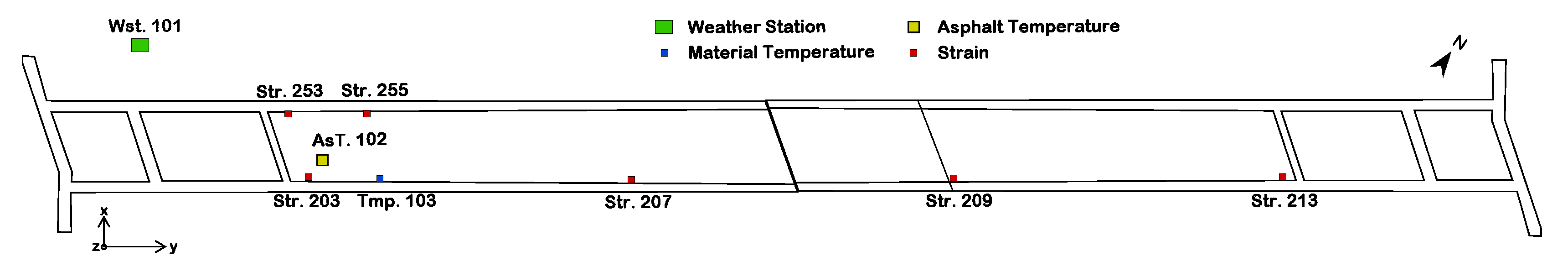}
    \caption{Vahrendorfer Stadtweg bridge from the west side (top left) and with three additional masses (top right). An overview of the sensor placement is found at the bottom.} 
    \label{fig:vs}
\end{figure}

The first case study is the prestressed concrete bridge ``Vahrendorfer Stadtweg'' as shown in Figure~\ref{fig:vs}~(top left). This structure spans the A7 freeway south of Hamburg, Germany, and includes a single lane designed for agricultural vehicles as well as a pedestrian walkway on its southeastern side. It was built in 1972, and measures 50 meters in length and 10 meters in width. Constructed with an open-frame design, the bridge features a box-girder cross-section. For the analysis presented here, measurements from six strain sensors (in the z-direction) were considered. As covariates, one material and one asphalt temperature sensor, as well as the relative humidity from the weather station located on the bridge, were used. The original sampling frequencies of 200~Hz and 10~Hz were downsampled (averaged) to one measurement every ten minutes, as referenced in \citep{Han.etal_2021}. The placement of the sensors is depicted in Figure~\ref{fig:vs}~(bottom). 

Load tests on the bridge were conducted from February 22nd to March 23rd, 2024. These tests used three large sand-filled bags weighing 680 kg, 740 kg, and 740 kg, positioned at the center of the bridge. Each additional weight was applied for ten days, increasing stepwise from 680 kg (scenario A) to 1,420 kg (scenario B) and 2,160 kg (scenario C). In Figure~\ref{fig:vs}~(top right), the maximum mass, which includes all three bags, is shown.
Subsequently, but not included in the presented analysis, the weights were relocated to the quarter point of the bridge, and after their removal, truck crossings were performed.
For more information regarding the bridge and the load tests, see \cite{Koehncke.etal_2026a}.\\


Data from July 2, 2023, to February 1, 2024, were used for training purposes (Phase I).
The conditional mean and covariance were estimated for temperature ranges from $-3^\circ$C to $28^\circ$C and humidity from $29.9\%$ to $100\%$ by using the five methods from above. Specifically:

\begin{itemize}
    \item[(i)] \textbf{Nadaraya-Watson:} The conditional mean and covariance were estimated using the Nadaraya-Watson kernel estimator as described in Section~\ref{sec:nwk} using the Euclidean norm. The bandwidths chosen via cross-validation led to apparent oversmoothing; therefore, reduced global bandwidths of $1.5$ and $2.5$ were chosen, respectively. \smallskip
    \item[(i)] \textbf{Random forest:} The conditional mean and covariance were estimated using a random forest approach as described in Section~\ref{sec:rf}. For the mean, the \texttt{R} package \texttt{randomForest}~\citep{Liaw.Wiener_2002} was used, and the covariance was estimated using the \texttt{R} package \texttt{CovRegRF}~\citep{Alakus.etal_2023} with 500 trees. \smallskip
    \item[(iii)] \textbf{Additive model:} Both the conditional mean and covariance were estimated using an additive model as described in Section~\ref{sec:add} employing the \texttt{R} package \texttt{mgcv}~\citep{Wood_2011,Wood_2017}, penalized (cubic) regression splines, and a quasi-likelihood family with an identity link and constant variance function. \smallskip
    \item[(iv)] \textbf{Additive model with interactions:} Both the conditional mean and covariance were estimated using an additive model with two-way interactions as described in Section~\ref{sec:add}. As before in Section~\ref{sec_validation_of_ana_method}, the \texttt{R} package \texttt{mgcv}~\citep{Wood_2011,Wood_2017} was used with penalized regression splines, tensor products for the interaction parts, and a quasi-likelihood family with identity link and constant variance. \smallskip
    \item[(v)] \textbf{Neural network:} The conditional mean and covariance functions were estimated using neural networks as described in Section~\ref{sec:nn} using the \texttt{Julia} package \texttt{Flux}~\citep{Bezanson.etal_2017}. For each strain sensor, the mean is modeled using a specific neural network with $2$ hidden layers of size $2q$ and $q$, respectively. For the variance and covariance, two different neural networks are used, each with $2$ hidden layers of size $3q$ and $4q$, respectively. \smallskip
\end{itemize}

\begin{figure}[htb!]
    \centering
    \includegraphics[width = .95\textwidth]{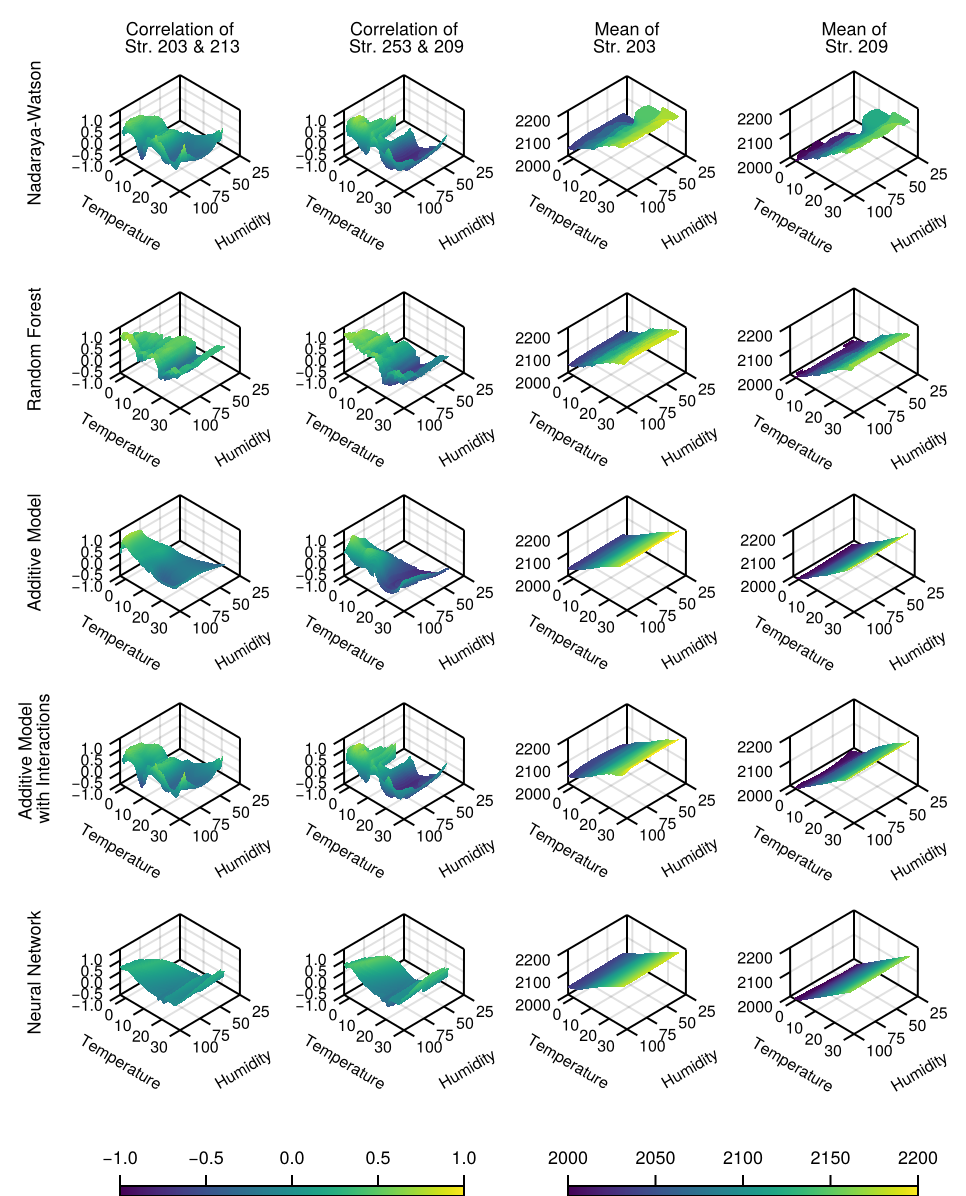}
    \caption{Conditional correlation (left columns) and mean (right columns) of the Vahrendorfer Stadtweg bridge given the material and asphalt temperature (x-axis) and the humidity (y-axis) using (top to bottom) the Nadaraya-Watson kernel estimator, a random forest, an additive model, an additive model with interactions, and a neural network.} 
    \label{fig:vs_ccn}
\end{figure}
Some of the resulting estimates are shown in Figure~\ref{fig:vs_ccn}.
The conditional mean of two strain sensors (203 and 209) as a function of the material and asphalt temperature, as well as the humidity, is shown in the two rightmost columns. Since a function of three input variables (material temperature, asphalt temperature, humidity) cannot be visualized in $\mathbb{R}^3$, the value of the function is drawn with material and asphalt temperature having the same value (the ``temperature'' axis). The plots clearly show a linear relationship with temperature and almost no influence of the humidity. The results for the other strain sensors (not shown) are very similar.

In the left two columns of Figure~\ref{fig:vs_ccn}, the conditional correlation is shown for two strain sensor pairs (203 \& 213, 209 \& 253). Here, the conditional correlations, instead of the covariances, are shown because they are easier to interpret due to the normalization to the $[-1,1]$ range. 
As before, the influence of the temperature(s) is much stronger than the influence of the humidity. Also, distinct differences between the methods are evident. The estimates produced by the additive model that includes interactions appear unstable and hardly interpretable, particularly at the boundaries, and closely resemble the results from the Nadaraya-Watson kernel estimator, which makes no structural assumptions. The estimates produced by the random forest are somewhat different but also wiggly, with a step-like pattern. If using a purely additive model, on the other hand, the depicted surfaces are much smoother, making it clear for strain sensors 203 and 213 (third row, first column), for instance, that the correlation is high at low temperatures and decreases at milder temperatures. The neural network also produces rather smooth estimates, but, like all methods, it seems to have problems at high temperatures, where training data is sparse.

\begin{figure}[h!]
    \centering
    \includegraphics[width=.9\textwidth]{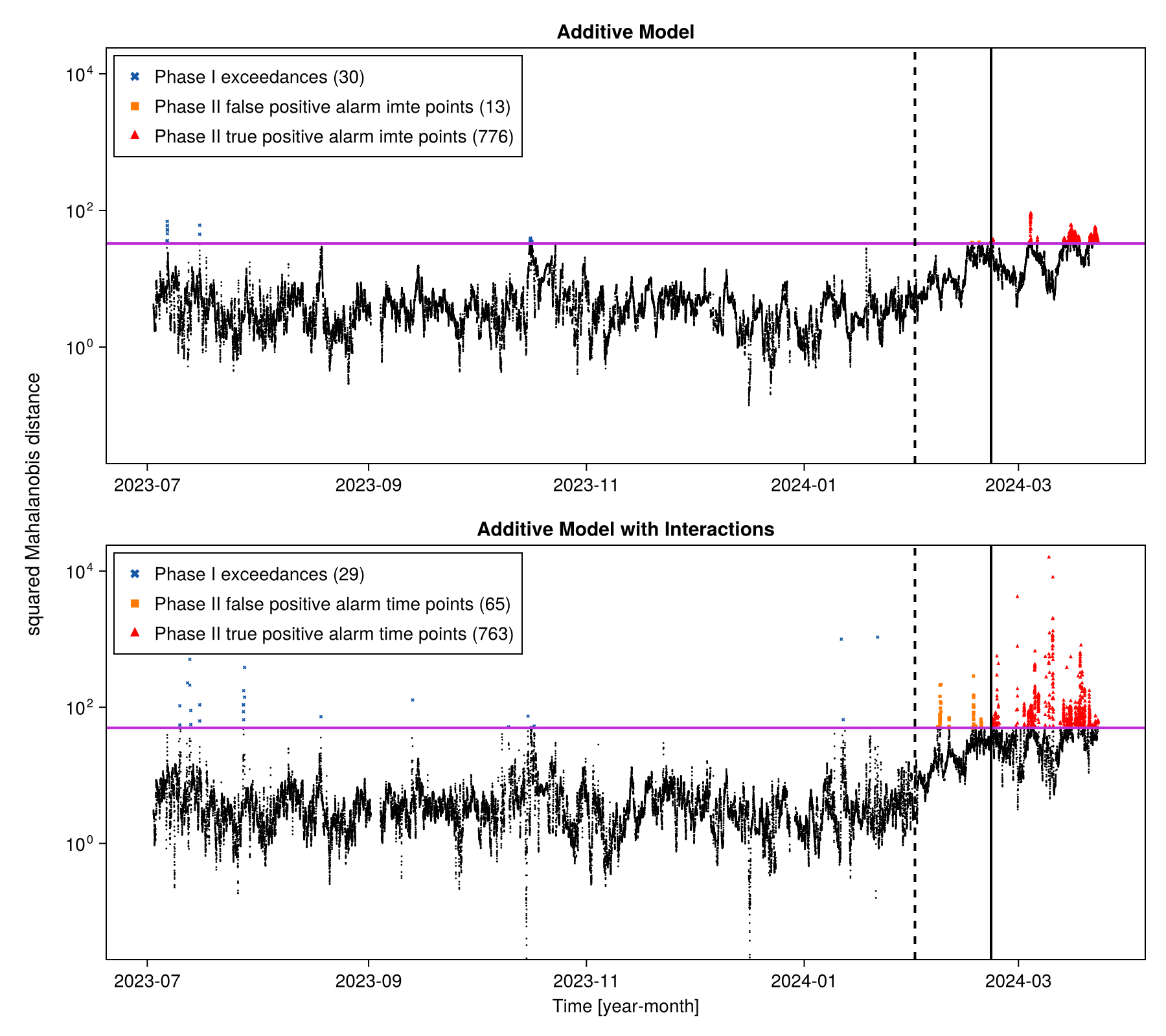}
    \caption{Squared conditional Mahalanobis distances for the additive model (top) and the additive model with interactions (bottom), shown on a logarithmic y-axis.  The horizontal purple line denotes the empirical control limit, defined as the \SI{99.9}{\percent} quantile of the Phase-I reference values. Threshold crossings are highlighted as Phase-I reference exceedances, Phase-IIa false-positive alarm time points, and Phase-IIb true-positive alarm time points. The dashed vertical line marks the end of Phase I, and the solid vertical line marks the start of the additional masses.}
    \label{fig:md}
\end{figure}
For monitoring purposes, the conditional Mahalanobis distance was calculated. Figure~\ref{fig:md} shows the results in terms of the squared CMD for the purely additive model (top) and the additive model with interactions (bottom). The comparison illustrates the difference between a structurally more restrictive model with additive covariate effects and a more flexible model that also allows covariate interactions. The dashed vertical line marks the end of Phase I, and the solid vertical line indicates the start of the additional masses. The horizontal purple line denotes the empirical control limit $h$, defined as the $99.9\%$ empirical quantile of the Phase-I training values. This corresponds to a nominal exceedance probability of $0.1\%$, or one exceedance per 1000 monitoring time points under an independent reference sequence. Since the CMD sequence is serially dependent, the resulting limit is interpreted as a marginal reference threshold. Threshold crossings in Phase I are shown as exceedances relative to the reference period. Threshold crossings in the in-control part of Phase II are counted as false-positive alarm time points and are colored orange. Threshold crossings after the start of the additional masses are counted as true-positive alarm time points and are colored in red.
Since the two models yield different stochastic behavior of the CMD sequence, the control limits differ: $h = 32.83$ for the purely additive model and $h = 49.73$ for the additive model with interactions.
If interactions among the three covariates are ignored, the results are smoother and generate substantially fewer false-positive alarm time points in the in-control part of Phase II. In contrast, the interaction model yields a higher control limit but still generates more Phase-II false-positive alarm time points, with 65 compared with 13 for the purely additive model. It also detects fewer true-positive alarm time points after the additional masses are introduced.
The first alarm with the purely additive model occurred 2 hours and 40 minutes earlier than for the additive model with interactions. This appears immediately after the additional masses are introduced and may therefore reflect the transient effect of the car with trailer crossing the bridge during this operation. A possible explanation for the increased number of false-positive alarm time points of the interaction model at the beginning of Phase II is that this period contains combinations of temperature and humidity values that were sparsely represented or not observed in the Phase-I training data. In such regions of the covariate space, estimates of both the conditional mean and the conditional covariance can become less stable. The purely additive model appears more robust in this setting, presumably because its structural restriction reduces variability in estimation in sparsely sampled regions of the covariate space.

\subsection{Railway Bridge KW51}\label{sec_kw51}

The second case study is the modal data of the railway bridge KW51~\citep{Maes.Lombaert_2021}. The KW51 railway bridge is a bowstring steel railway bridge measuring 115~meters in length and 12.4~meters in width with two curved electrified tracks. It is located between Leuven and Brussels, Belgium. The bridge was monitored between October 2nd, 2018, and January 15th, 2020, with a retrofit period from May 15th to September 27th, 2019. The data prior to the retrofit are used as in-control data. Because some measurements of material temperature and relative humidity are missing during this period, only four months of data were used for model training here. The modal data was calculated hourly by \cite{Maes.Lombaert_2021} and provided online~\citep{Maes.Lombaert_2020}. From the 14 tracked eigenfrequencies, eight were used for analysis here (Modes 3, 5, 6, 9, 10, 12, 13, and 14). The others were excluded because they were not sufficiently excited and could not be tracked sufficiently. The missing data for the remaining eight modes were filled using linear interpolation as in previous studies~\citep{Maes.Lombaert_2021}. 

As also seen in previous studies~\citep{Maes.Lombaert_2021,Neumann.etal_2025b}, eigenfrequencies tend to be higher at colder temperatures and lower at warmer temperatures (above $2^\circ$C). Furthermore, variances are much larger at cold temperatures. This results in substantially increased false alarm rates at low temperatures if using standard approaches for anomaly detection; see, e.g., \cite{Neumann.etal_2025b}. As a consequence, confounding influences should be eliminated in both the mean and the (co-)variances. To do so, the conditional mean and covariances were estimated as in Section~\ref{sec_vs}. According to cross-validation, the optimal bandwidth for the Nadaraya-Watson kernel estimator should be large; hence, a global bandwidth of 5.5 was used. Due to the relatively simple relationship between the natural frequencies and temperature/humidity, the conditional mean could be estimated using a rather simple neural network with $1$ hidden layer of size $3q$. For the variance and covariance, two different neural networks were used, each with $2$ hidden layers of size $2q$ and $3q$, respectively. 

\begin{figure}[htb!]
    \centering
    \includegraphics[width=\textwidth]{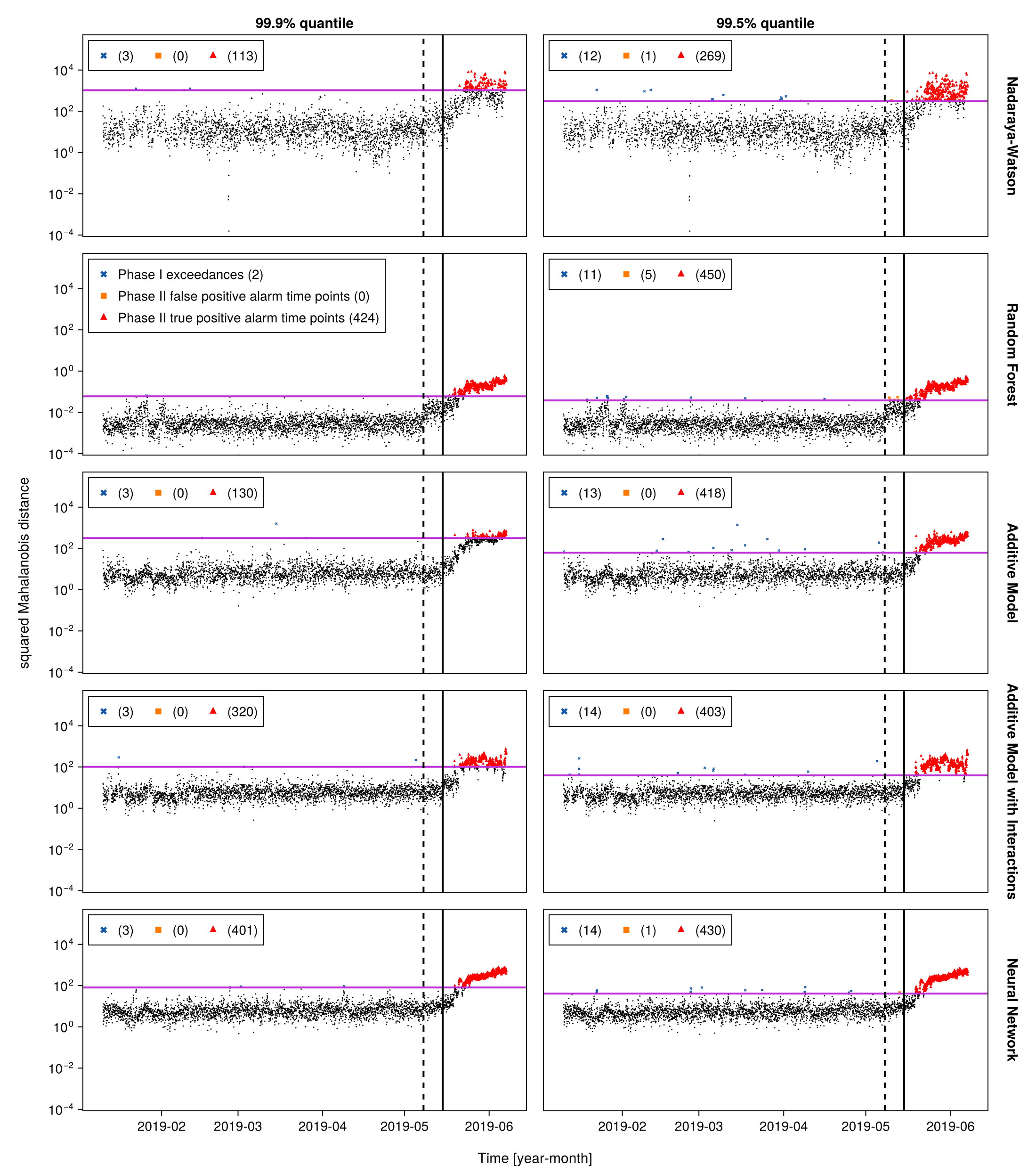}
    \caption{Squared conditional Mahalanobis distances obtained with the five methods for the KW51 bridge eigenfrequency data, shown on a logarithmic y-axis. The horizontal purple line denotes the empirical control limit, defined as the \SI{99.9}{\percent} (left) and \SI{99.5}{\percent} (right) quantile of the Phase-I reference values. The dashed vertical line marks the start of Phase II, and the solid vertical line marks the start of the retrofit, which is treated as the sustained out-of-control condition. Threshold crossings are highlighted as Phase-I exceedances (blue), Phase-II false-positive alarm time points (orange), and Phase-II true-positive alarm time points (red).}
    \label{fig:md_kw51}
\end{figure}

Figure~\ref{fig:md_kw51} shows the (squared) conditional Mahalanobis distance for the five methods considered: Nadaraya-Watson (first row), random forest (second row), purely additive model (third row), additive model with interactions (fourth row), and neural network (fifth row). The columns represent two different control limits (horizontal purple lines): the 99.9\% (left) and 99.5\% (right) empirical quantiles from the Phase I data, which is the data to the left of the dashed vertical line. 
The resulting exceedances in Phase I are colored blue, and the false-positive (in-control) and true-positive (out-of-control) alarms in Phase II are colored orange and red, respectively. The start of the retrofit period is indicated by the solid vertical line. 
Overall, most methods show good monitoring performance. The conditional
Mahalanobis distance largely mitigates the accumulation of false-positive alarms during cold-temperature periods, as observed with standard EOV removal methods~\citep{Dervilis.etal_2015,Maes.etal_2022,Neumann.etal_2025b}; only the random-forest approach, combined with the notably lower control limit, shows a small cluster of false-positive alarm time points around the end of January 2019. After the onset of the retrofit, the Nadaraya--Watson estimator performs worse than the other methods, with substantially lower post-retrofit alarm coverage. This may be due to the lower stability of fully nonparametric estimates of the conditional mean and covariance in this application. Among the best-performing methods, the random forest is the most sensitive: it produces the highest number of post-retrofit true-positive alarm time points and the earliest first alarm, about one to three days earlier than the other methods. This increased sensitivity, however, may come at the cost of more Phase-II false-positive alarm time points before the retrofit (as seen with the 99.5\% quantile in the right column). The additive models and the neural network appear more conservative, with somewhat lower post-retrofit alarm coverage but potentially fewer false-positive alarm time points. In summary, particularly the random forest, the neural network, and the additive model with interaction terms seem to provide good compromises between model complexity/flexibility and robustness.

\section{Summary, Discussion, and Conclusion}
\label{sec_conclusion}

The main contributions of this paper are novel approaches for estimating the conditional covariance of system outputs in SHM under multivariate, varying environmental conditions. Two existing methods, nonparametric kernel-based estimates and a random forest estimator, and two newly proposed supervised-learning approaches employing semiparametric additive modeling or neural networks, were discussed. If using the additive modeling approach, different specifications are possible. Particularly, interactions between the covariates can be included, or a purely additive model structure can be chosen. 
This fills a gap in the literature, which has so far focused on univariate confounding influences. 
In particular, new methods that may replace existing nonparametric, kernel-based approaches allow the use of multiple covariates, as they are less susceptible to the ``curse of dimensionality''. Furthermore, the resulting conditional covariances, together with estimates of the conditional mean, can be used to remove multivariate environmental or other confounding influences from the data to be used for (structural health) monitoring purposes. Several approaches to do so, such as the conditional Mahalanobis distance, were described in the paper.

A comparative Monte Carlo simulation study was conducted to compare the performance of various methods for estimating conditional covariances. Although all methods reproduced the underlying correlation used for data generation, some differences were observed, particularly between the random forest, which combines piecewise constant regression trees, and the other methods that produce smooth surfaces. Regarding estimation accuracy, as measured by root mean square error (RMSE), the additive models, with and without interaction, performed best, followed by the neural network. The random forest apparently had problems fitting the underlying, truly smooth regression functions. 

As also shown in the paper, the developed methods can be applied to both static response measurements (strain) and damage-sensitive features (eigenfrequencies). This was demonstrated using two benchmark datasets: load test data from the Vahrendorfer Stadtweg bridge and modal data from the railway bridge KW51. 
In the first case study, it was shown, using the discussed approaches, that the influence of humidity on both the mean and the covariance of the strain sensors is much weaker than that of temperature. Regarding anomaly detection (i.e., detection of the added masses) using conditional Mahalanobis distances, the purely additive model produced the most reliable results, presumably due to its robustness to data sparsity in the predictor space.

Using the conditional Mahalanobis distance for the modal data of the railway bridge KW51, the influence of covariates could be eliminated by using any of the discussed approaches for covariance estimation. Without fully removing confounding influences, previous studies have shown that the MD spiked at colder temperatures, where the eigenfrequencies are higher and exhibit higher variance and correlation. Using the conditional MD version, these effects could be eliminated while still detecting the structural changes caused by the retrofit. 

The case of the Vahrendorfer Stadtweg bridge, however, also showed limitations of the proposed approaches. Even more than in settings with one-dimensional covariates, it is crucial to have rich training data available that spans the space of potential predictor values. This is particularly true for methods that can capture complex relationships between the covariates and the system outputs, such as nonlinear interactions. If, for example, certain combinations of covariate values are not included in the training dataset, those methods will perform poorly in those regions of the predictor space, increasing the risk of false alarms.
%
In conclusion, the performance of the approaches discussed depends on both the true underlying relationships between the covariates and the system outputs and the amount and quality of the available training data. As illustrated in the paper, an additive model that can fit nonlinear associations while still imposing some (additive) structure on the modeled covariate effects is often a good compromise between flexibility and robustness.

\section*{Software}
\label{sec:Software}

The data analysis was performed using the statistical software \texttt{Julia} \citep{Bezanson.etal_2017} and \texttt{R} \citep{R_2025}.
Source codes for loading and preprocessing the KW51 data~\citep{Maes.Lombaert_2020}, implementing the five different models from Section~\ref{sec_nonoparm_est_cond_cov} and the conditional Mahalanobis distance from Section~\ref{sec_SHMmethods} for the reproducing the results from Section~\ref{sec_kw51} are available on
GitHub at \url{https://github.com/neumannLizzie/FeatureReconstructionSHM/}. 

\section*{Data availability}\label
{sec:data_availability}

The load test data of the Vahrendorfer Stadtweg bridge is available on Zenodo~\citep{Koehncke.etal_2026b} and the full data set from the corresponding author of \cite{Koehncke.etal_2026a} upon reasonable request.
The data of the railway bridge KW51 is available from Zenodo~\citep{Maes.Lombaert_2020}.

\section*{Acknowledgements}

This research paper out of the project `SHM -- Digitalisierung und Überwachung von Infrastrukturbauwerken' is funded by dtec.bw -- Digitalization and Technology Research Center of the Bundeswehr, which we gratefully acknowledge. 
Computational resources (ISCC) have been provided by the project hpc.bw, funded by dtec.bw – Digitalization and Technology Research Center of the Bundeswehr. dtec.bw is funded by the European Union – NextGenerationEU. 



\end{document}